# Maximum Entropy for Collaborative Filtering

C. Lawrence Zitnick
Microsoft Research
Redmond, WA 98052

Takeo Kanade
Robotics Institute
Carnegie Mellon University
Pittsburgh, PA 15213

## Abstract

Within the task of collaborative filtering two challenges for computing conditional probabilities exist. First, the amount of training data available is typically sparse with respect to the size of the domain. Thus, support for higher-order interactions is generally not present. Second, the variables that we are conditioning upon vary for each query. That is, users label different variables during each query. For this reason, there is no consistent input to output mapping. To address these problems we purpose a maximum entropy approach using a non-standard measure of entropy. This approach can be simplified to solving a set of linear equations that can be efficiently solved.

## 1 Introduction

The goal of collaborative filtering is to predict the preferences of a user given the preferences of others. For example, given a set of movies a user likes, we would like to predict what movies the user might also like given the opinions of others. The application of collaborative filtering assumes that user preferences are correlated, i. e. there exists groups of people with similar preferences.

We formulate collaborative filtering as computing the probability of a user desiring an item conditioning upon their past known preferences and the preferences of others. Typically, collaborative filtering domains have thousands of variables with limited training data, or past user data. For one example within this paper, we have 1,600 variables with 5,000 past users each labeling on average only 46 variables.

Numerous algorithms and domains for collaborative filtering have been proposed. Some earlier domains include (Goldberg et al. 1992) with email and (Resnick et al. 1994) with netnews. A comparison of earlier works can be found in (Breese et al. 1998). More recently, SVD decomposition (Billsus and Pazzani 1998), dependency networks (Heckerman et al. 2000) and graphic models (Jin et al. 2003) have been proposed for use in collaborative filtering.

We propose a method based on maximum entropy. That is, we attempt to enforce a set of constraints on a distribution while maximizing its entropy. The set of constraints will take the form of conditional probabilities that can be accurately computed from the data. These conditional probabilities will typically be of low-order and will not fully constrain the probability distribution. We will propose an extension to the algorithm that allows the constraints to be partially enforced based on our confidence in them. Thus, new constraints can be added to reduce the errors in our system due to bias, while not increasing the errors due to variance.

For problems in which the input to output mapping isn't fixed, maximizing Shannon's measure of entropy is too computationally expensive. We propose using a alternative measure of entropy which lies within the family of Rényi's entropies. Using this measure we can reduce the algorithm to solving a set of linear equations. While we address the problem of collaborative filtering within this paper, the results and algorithm may be applied to more general problems involving inference.

Previously, maximum entropy has been applied to collaborative filtering by (Pavlov and Pennock 2002.) Their task was to predict which documents a user might visit next given their past viewing history. Since all documents have a known value of viewed or not viewed, a fixed input to output mapping exists, similar to (Shani et al. 2002) and the natural language processing task (Rosenfeld 1994.) For this reason, the use of conventional entropy techniques was feasible. For problems examined in this paper, the set of unknown variables varies with each query. For example, as input each user may label a different subset of movies as liked or not liked. This makes the use of these same techniques computationally expensive.

Before we describe our algorithm, we will discuss notation and maximum entropy methods in sections 2 and 3. In section 4 we will describe the algorithm followed by some collaborative filtering results in section 5.

## 2　Notation

Our world is described by a set of binary variables $X = \{X_1, \ldots, X_a\}$ with a corresponding set of values $x = \{x_1, \ldots, x_a\}$ with $x_i \in \{0, 1\}$ for all $i \in a$. Each variable represents an item such as a specific book, movie or web page. A variable is assigned the value of one if the user labels the item as desired, and a value of zero if it is not. We are given an $m \times a$ set of training data $T$, where $m$ is the number of entries or people in the training set. We will refer to the value of the $i$th variable in the $j$th entry as $t_{j,i}$, and the entire set of variable values in the $j$th entry as $t_j$. From the training set we can compute the empirical distribution $\tilde{P}$ by:

$$\tilde{P}(X = x) = \frac{1}{m} \sum_{j \in m} \delta(x, t_j) \qquad (1)$$

where

$$\delta(x, t_j) = \left\{ \begin{array}{ll} 1 & x = t_j \\ 0 & x \neq t_j \end{array} \right\} \qquad (2)$$

At any particular time, some of the variables will be observed while others are not, i. e. the user will label some subset of the variables. The evidence variables representing the set of observed or labeled variables will form the set $X_E$. The hidden variables that are unobserved will form the set $X_H = X - X_E$. The number of variables in $X_E$ is denoted by $e$ and the number of variables in $X_H$ is denoted as $h$, $e + h = a$. It is our goal to compute the value of the conditional probabilities $P(X_i \mid X_E)$ for all $X_i \in X_H$.

The simplest approach to collaborative filtering is to compute the conditional probability of $P(X_i = 1 \mid x_E)$ from the empirical distribution:

$$\tilde{P}(X_i = 1 | x_E) = \frac{\sum_{j \in m} t_{j,i} \beta_j(x_E)}{\sum_{j \in m} \beta_j(x_E)} \qquad (3)$$

where

$$\beta_j(x_E) = \left\{ \begin{array}{ll} 1 & t_{j,k} = x_k \text{ for all } X_k \in X_E \\ 0 & \text{otherwise} \end{array} \right\} \qquad (4)$$

We abbreviate $P(X_i = 1 \mid X_E = x_E)$ as $P(X_i = 1 \mid x_E)$. Unfortunately, for many $x_E$ no examples will exist in the user data set. In these cases, a common approach is to look for partial matches or to create some measure of similarity between users or queries. We take a different approach to collaborative filtering in that we attempt to find relations between variables, instead of users.

| User | Romance 1 | Romance 2 | Action 1 | Action 2 |
|---|---|---|---|---|
| 1 | 0 | 1 | 0 | 0 |
| 2 | 1 | 1 | 0 | 0 |
| 3 | 0 | 0 | 0 | 1 |
| 4 | 0 | 0 | 1 | 1 |
| 5 | 0 | 0 | 1 | 1 |
| 6 | 1 | 1 | 1 | 0 |

Table 1: Training data for sample problem.

### 2.1　Sample Problem

To help in clarifying many of the points within this paper we will use a small sample problem. Our problem will consist of four variables $X = \{X_{r1}, X_{r2}, X_{a1}, X_{a2}\}$ representing two romance movies $X_{r1}$ and $X_{r2}$ and two action movies $X_{a1}$ and $X_{a2}$. We are given six user entries, $m = 6$, as shown in Table 1.

Therefore, if a new user labels the first romance movie as being desired $X_{r1} = 1$ then $X_E = \{X_{r1}\}$ and $X_H = \{X_{r2}, X_{a1}, X_{a2}\}$ with:

$$\tilde{P}(X_{r2} = 1 | x_E) = 1.00 \qquad (5)$$
$$\tilde{P}(X_{a1} = 1 | x_E) = 0.50 \qquad (6)$$
$$\tilde{P}(X_{a2} = 1 | x_E) = 0.00 \qquad (7)$$

## 3　Maximum Entropy

As the number of variables we're conditioning upon increases, the accuracy of $\tilde{P}(X_i = 1 | x_E)$ to estimate $P(X_i = 1 | x_E)$ decreases. Within our data sets, it's not uncommon for a user to show preferences for just three variables for which no previous user with similar tastes exists.

While we may not be able to accurately compute $P(X_i = 1 | x_E)$ directly from the data, we may be able to accurately compute $P(X_i = 1 | x_S)$ for some subset $x_S \subset x_E$. We will use these subsets of $x_E$ to help form a set of constraints on our conditional distribution.

Before we define our constraints, let us first create a set of binary indicator functions $F$. These functions consist of a logical statement on some subset of the variables $X_E$. More specifically they correspond to sets of variable values $x_S$ for which we believe $\tilde{P}(X_i = 1 | x_S)$ closely approximates the true distribution.

For example, if our user labels movies $X_{r2}$ and $X_{a1}$ as desired, we'd like to compute $P(X_{r1} = 1 | X_{r2} =$

$1, X_{a1} = 1$). Unfortunately, computing this value using the empirical distribution may be inaccurate since only one entry in the training set has $X_{r2} = 1$ and $X_{a1} = 1$. However, we can compute the values of $P(X_{r1} = 1)$, $P(X_{r1} = 1|X_{r2} = 1)$ and $P(X_{r1} = 1|X_{a1} = 1)$ more reliably. Thus we'd create three indictor functions $f_0(x_E)$, $f_1(x_E)$ and $f_2(x_E)$ equal to:

$$f_0(x_E) = 1 \quad (8)$$

$$f_1(x_E) = \left\{ \begin{array}{ll} 1 & x_{r2} = 1 \\ 0 & \text{otherwise} \end{array} \right\} \quad (9)$$

$$f_2(x_E) = \left\{ \begin{array}{ll} 1 & x_{a1} = 1 \\ 0 & \text{otherwise} \end{array} \right\} \quad (10)$$

The set of functions $F = \{f_0, \ldots, f_c\}$ and the empirical frequencies $\tilde{P}(X_i \mid f_k)$ form a set of constraints on our computed distribution $P$:

$$\tilde{P}(X_i = 1, f_k) = \sum_{x_E} \tilde{P}(x_E) P(X_i = 1 \mid x_E) f_k(x_E) \quad (11)$$

The set of constraints will not fully constrain the values of the joint distribution, with $c \ll 2^e$ typically.

When computing our estimate of the true conditional distribution, we'd like to enforce the constraints while not introducing other bias. In general, the uniform distribution is typically regarded as the most unbias distribution. For example, if we know nothing about a coin, it is usually assumed to be fair. Similarly, if we'd want to find the most unbias distribution given the constraints, we should find the distribution that lies closest to the uniform distribution that also obeys the constraints.

The distance between any distribution and the uniform distribution may be measured by their relative entropy. Entropy is a measure of the average amount of information needed to describe the variables at any particular time. Since the uniform distribution is the distribution with highest entropy, we may simplify our task to just maximizing the entropy of the constrained distribution (Jaynes 1957; Kullback 1959).

Maximum entropy methods have the advantage that they choose the least committal solution to a problem given the constraints, i. e. assume independence until proven otherwise. Similarly, Jaynes (Jaynes 1990) has said

> Maximum entropy agrees with everything that is known, but carefully avoids anything that is unknown.

### 3.1 Shannon's Entropy

One of the first measures of entropy and still most popular is that of Shannon (Shannon and Weaver 1963). Shannon developed three properties for a measure of information in a communication stream. Later research has applied his measure to a wide range of applications including but not limited to spectral analysis (Burg 1967), language modeling (Rosenfeld 1994) and economics (Golan et al. 1996). Shannon constructed his measure $H$ so that it satisfied the following properties for all $p_i$ within the estimated joint probability distribution $P$:

1. $H$ is a continuous positive function.

2. If all $p_i$ are equal, $p_i = \frac{1}{n}$, then $H$ should be a monotonic increasing function of $n$.

3. For all $n \geq 2$, $H(p_1, \ldots, p_n) = H(p_1 + p_2, p_3, \ldots, p_n) + (p_1 + p_2) H(\frac{p_1}{p_1+p_2}, \frac{p_2}{p_1+p_2})$.

Shannon showed the only function that satisfied these properties is:

$$H(P) = -\sum_i p_i \log(p_i) \quad (12)$$

Since we are concerned with conditional probabilities, we will maximize the conditional form of Shannon's entropy, i. e. we will maximize the expectation of the entropy given $x_E$:

$$H(X_i \mid X_E) = -\sum_{x_E} P(x_E) \sum_{x_i} P(x_i \mid x_E) \log(P(x_i \mid x_E)) \quad (13)$$

Using the Lagrangian function the distribution $P$ that satisfies our constraints (11) while maximizing the entropy function (13) has the following form:

$$P(X_i = 1 \mid x_E) = \prod_i \mu_i^{f_i(x_E)} \quad (14)$$

The set of Lagrangian multipliers $\mu_i$ can be computed using the "Generalized Iterative Scaling" algorithm (GIS) (Darroch and Ratcliff 1972) or by using some variant of gradient descent. Unfortunately, these algorithms can be quite computationally expensive. Theoretically, the algorithms require summing over the entire set of possible values for $X_E$. Thus the running time is $O(2^e)$. Previous work has modified equation (13) to use $\tilde{P}(x_E)$ instead of $P(x_E)$. However, this still requires searching through the entire user data set.

Further complicating the problem, is the fact that the set of evidence variables will vary for each query made to the system. Therefore the parameters $\mu_i$ cannot be pre-computed prior to a query. For these reasons, the use of Shannon's entropy for computed conditional probabilities is too computationally expensive for use in collaborative filtering.

### 3.2 Rényi's Entropy

About ten years after Shannon introduced his measure of entropy, a mathematician from Hungary named Rényi (Rényi 1976a, 1976b, 1976c) generalized his work. Rényi relaxed Shannon's third property for measures of entropy $H_\alpha$ as follows:

For two independent distributions $P$ and $\acute{P}$:
$$H_\alpha(P\acute{P}) = H_\alpha(P) + H_\alpha(\acute{P}) \quad (15)$$

Rényi found that the following family of functions satisfies Shannon's first two properties and his generalization of the third (Rényi referred to his family of information measures as $I_\alpha$):

$$H_\alpha(P) = \frac{1}{1-\alpha} \log\left(\sum_i p_i^\alpha\right) \text{ for } \alpha > 0 \quad (16)$$

As $\alpha$ approaches one, equation (16) reverts back to Shannon's equation (12), that is:

$$\lim_{\alpha \to 1} H_\alpha(P) = H(P) \quad (17)$$

Of particular interest to us is the case when $\alpha$ is equal to two. This measure has been called Rényi's Quadratic Entropy (RQE):

$$H_2 = -\log\left(\sum_i p_i^2\right) \quad (18)$$

Since we are only concerned with maximizing the entropy we can drop the log and minus sign from the equation which results in minimizing:

$$\sum_i p_i^2 \quad (19)$$

Once again, since we are concerned with computing conditional probabilities, we will compute the expected value of (19) given the values $x_E$:

$$\sum_{x_E} P(x_E) \sum_{x_i} P(x_i \mid x_E)^2 \quad (20)$$

We will approximate the above equation using the empirical frequency $\tilde{P}(x_E)$ instead of $P(x_E)$:

$$\sum_{x_E} \tilde{P}(x_E) \sum_{x_i} P(x_i \mid x_E)^2 \quad (21)$$

Using the Lagrangian function we find our probability estimates $y_i$ for $P(X_i = 1 \mid x_E)$ that minimize (21) while enforcing the constraints (11) take the following form:

$$y_i = \sum_j \lambda_{i,j} f_j(x_E) \quad (22)$$

where $\lambda_{i,j}$ are the Lagrangian multipliers. We refer to the values $y_i \approx P(X_i = 1 \mid x_E)$ as probability estimates and not as probabilities themselves since it is possible that their values may not lie between 0 and 1, as explained in section 4.2.

The measure (19) may also be interpreted as the Brier score (Grünwald and Dawid 2002).

## 4 Maximizing Rényi's Quadratic Entropy without Bounds

From the preceding section we learned that our probability estimates $y_i$ of $P(X_i = 1 \mid x_E)$ can be computed from a weighted linear sum (22) of the functions $F$. Thus, for each query we need to compute the weights $\lambda_{i,j}$. Combining the constraint function (11) and (22) we find the following:

$$\tilde{P}(X_i = 1, f_j) = \\ \sum_{x_E} \tilde{P}(x_E) \sum_k \lambda_{i,k} f_k(x_E) f_j(x_E) \quad (23)$$

Using the fact that:

$$\tilde{P}(f_j, f_k) = \sum_{x_E} \tilde{P}(x_E) f_j(x_E) f_k(x_E) \quad (24)$$

and rearranging the summations we find:

$$\tilde{P}(X_i = 1, f_j) = \sum_k \lambda_{i,k} \tilde{P}(f_j, f_k) \quad (25)$$

We can create a list of functions for all possible user queries, and pre-compute the pairwise empirical frequencies $\tilde{P}(f_j, f_k)$ for all pairs of functions. Thus, we eliminate the need to sum over all possible values of $X_E$ for each query.

We may rewrite the set of equations (25) for all $j \in c$ in matrix notation. If:

$$\mathbf{p}_i = \begin{bmatrix} \tilde{P}(X_i = 1|f_0) \\ \vdots \\ \tilde{P}(X_i = 1|f_c) \end{bmatrix} \quad (26)$$

and

$$\mathbf{P} = \begin{bmatrix} \tilde{P}(f_0|f_0) & \tilde{P}(f_0|f_1) & \cdots & \tilde{P}(f_0|f_c) \\ \vdots & \vdots & \ddots & \vdots \\ \tilde{P}(f_c|f_0) & \tilde{P}(f_c|f_1) & \cdots & \tilde{P}(f_c|f_c) \end{bmatrix} \quad (27)$$

then

$$\mathbf{p}_i^\mathbf{T} = \lambda_i \mathbf{P} \quad (28)$$

Therefore we can compute our weights $\lambda_{i,j}$ from:

$$\lambda_i = \mathbf{p}_i^{\mathbf{T}}\mathbf{P}^{-1} \qquad (29)$$

Since inverting a $n \times n$ matrix requires $O(n^3)$ computation, the running time for our algorithm is $O(c^3 + ca)$. If $c$ is significantly smaller than $a$, as is typically the case, then the algorithm runs in approximately linear time $O(a)$ with respect to $a$.

### 4.1 Example

Given our sample problem described in section 2.1, and using the indictor functions (8), (9) and (10), we can attempt to solve for $P(X_{r1} = 1|X_{r2} = 1, X_{a1} = 1)$. Using the data in table 1 we find the vector $\mathbf{p}_i$ and matrix $\mathbf{P}$ to be:

$$\mathbf{p}_i = \begin{bmatrix} 0.33 \\ 0.67 \\ 0.33 \end{bmatrix} \qquad \mathbf{P} = \begin{bmatrix} 1.00 & 1.00 & 1.00 \\ 0.50 & 1.00 & 0.33 \\ 0.50 & 0.33 & 1.00 \end{bmatrix}$$

Solving for the $\lambda$s using (29) we find: $\lambda_{r1,0} = -0.167$, $\lambda_{r1,1} = 0.750$ and $\lambda_{r1,2} = 0.250$. Therefore using (22), our estimate $y_{r1}$ of $P(X_{r1} = 1|X_{r2} = 1, X_{a1} = 1)$ is equal to 0.833, which is most likely a better estimate than the value of 1.0 found using the empirical frequency.

### 4.2 Bounding Constraints

It is possible that our probability estimates $y_i$ might lie outside the range of $[0, 1]$, since we didn't enforce the inequality constraints that $y_i \geq 0$ and $y_i \leq 1$. Unfortunately, these bounding constraints cannot be enforced using our method of Lagrangian multipliers, and is the reason we believe this method hasn't been used previously (Jaynes 1957.) If these bounding constraints are enforced, the computational advantages of using Rényi's quadratic entropy are lost.

For many applications such as collaborative filtering, in which items need to be ranked or only relative values are needed, the errors associated with ignored these constraints can be acceptable. This is especially true when the computational advantages are also considered.

### 4.3 Constraint Confidence

Previously, we assumed that each constraint value was known exactly. In real world problems this is not the case. For some constraints a large amount of data will be available and thus we can be confident of its value. However, many constraints will have less supporting data. We may not want to fully enforce a constraint that has a value for which we're not confident.

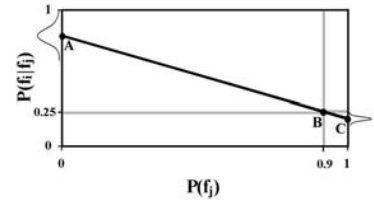

Figure 1: The linear relationship between $P(f_i|\neg f_j) \to A$, $P(f_i) \to B$ and $P(f_i|f_j) \to C$. As point $B$ approaches point $C$, the variance of $C$ decreases.

Until this point, the value of the constraint on $X_i$ given $f_j$ was set equal to $\tilde{P}(X_i = 1|f_j)$. Depending on the amount of user data available, the approximation of the true distribution using $\tilde{P}(X_i = 1|f_j)$ may not be accurate. Instead of assigning constraint values equal to $\tilde{P}(X_i = 1|f_j)$, we will compute a new set of constraint values $\mathbf{C}_{i,j}$ that takes into account our confidence in the value of the empirical frequency. The matrix $\mathbf{C}$ of constraint values can be used in place of $\mathbf{P}$ in the equations used for computing the weights $\lambda_{i,j}$.

The constraint values will be computed using a linear combination of $\tilde{P}(X_i = 1|f_j)$ and $\tilde{P}(X_i = 1)$:

$$\mathbf{C}_{i,j} = c_j \tilde{P}(X_i = 1|f_j) + (1.0 - c_j)\tilde{P}(X_i = 1) \quad (30)$$

The value $c_j \in [0, 1]$ is called the confidence value for the function $f_j$. If we have no confidence in the function $f_j$ we set $c_j = 0$ and $\mathbf{C}_{i,j} = \tilde{P}(X_i = 1)$. This results in the constraint values being equivalent to those for $\mathbf{C}_{i,0}$ since $\mathbf{C}_{i,0} = \tilde{P}(X_i = 1|f_0) = \tilde{P}(X_i = 1)$. Thus the constraint $\mathbf{C}_{i,j}$ will add no additional information, and the output will not depend on $f_j$. The closer the value of $c_j$ is to one, the greater the impact of $f_j$ on the outputs.

Assuming the empirical frequencies $\tilde{P}(f_i|f_j)$ have normal distributions, the variance of $\tilde{P}(f_i|f_j)$ decreases relative to $\frac{1}{m\tilde{P}(f_j)}$. Therefore, the error of $\tilde{P}(f_i|f_j)$ may be much larger than $\tilde{P}(f_i)$ or $\tilde{P}(f_j)$, if $\tilde{P}(f_j)$ is small.

If $P(f_j)$ is close to zero then we may not be able to accurately compute $P(f_i|f_j)$ since we will not observe $f_j = 1$ with high enough frequency. The same holds for $P(\neg f_j)$ and $P(f_i|\neg f_j)$. Consider figure 1, if we assume the values $P(f_i)$ and $P(f_j)$ are known then point $B$ is fixed. The values of $P(f_i|\neg f_j)$ and $P(f_i|f_j)$, corresponding to points $A$ and $C$ respectively, are related as follows:

$$\frac{P(f_i) - P(f_i|f_j)P(f_j)}{P(\neg f_j)} = P(f_i|\neg f_j) \qquad (31)$$

That is, the three points $A$, $B$ and $C$ lie along a line. Since point $B$ is fixed, we are essentially trying to find the slope of the line in figure 1. The variance of the slope varies relative to $\frac{1}{mP(f_j)P(\neg f_j)}$.

If we have no confidence in our constraint, then $P(f_i|f_j) = P(f_i|\neg f_j) = P(f_i)$. Thus, there will be zero slope. We may interpret this as assuming a prior distribution around the zero slope. We will assume the prior distribution is a normal distribution with a variance of $\sigma_0^2$ and mean 0. We can convolve this with the distribution obtained from computing the slope of the empirical frequencies. If we determine the variance of our computed slope to be $\sigma_j^2 = \frac{\varphi}{mP(f_j)P(\neg f_j)}$, then the mean or expected value of the two combined distributions will equal:

$$\frac{\sigma_0^2}{\frac{\varphi}{mP(f_j)P(\neg f_j)} + \sigma_0^2} \left( \tilde{P}(f_i|f_j) - \tilde{P}(f_i|\neg f_j) \right) \qquad (32)$$

Using both 31 and 32 we find the confidence value $c_j$ is equal to:

$$c_j = \frac{\sigma_0^2}{\frac{\varphi}{mP(f_j)P(\neg f_j)} + \sigma_0^2} \qquad (33)$$

All that remains for computing $c_j$ is setting the values of $\sigma_0^2$ and $\varphi$. Since $\varphi$ is a function of unknown values, namely $P(f_i, f_j)$ and $P(f_i, \neg f_j)$, we will set its value to a ratio of $\sigma_0^2$. Using $\varphi = r\sigma_0^2$ equation (33) reduces to:

$$c_j = \frac{mP(f_j)P(\neg f_j)}{r + mP(f_j)P(\neg f_j)} \qquad (34)$$

The value of $r$ may be set to a wide range of values depending on how much we'd like to bias the network's results towards the prior probabilities. In practice, $r$ is typically set between 5 and 100.

## 5 Results

We test our algorithm on two collaborative filtering data sets, MSWeb and EachMovie.

### 5.1 Web Browsing Behavior

Our first collaborative filtering task is predicting which web pages a user will visit given their previous browsing history. Our database consists of 32,711 training cases, where each case is a list of web pages an individual user visited. The testing database consists of 5,000 cases. The data sets are supplied courtesy of Microsoft Corporation from user logs generated during one day in 1996.

To measure the accuracy of our results we will use the same error metric as described in (Breese et al. 1998; Heckerman et al. 2000). For each test case the web pages visited will be split randomly into input and measurement sets. The input set will be given to our algorithm as a set of evidence values to compute the probability of the user visiting the other web pages. The web pages are then ranked based on their computed probabilities. If $K_i$ is the number of items in the measurement set, $R_i$ is the number of items on the recommendation list and $M$ is the total number of test cases, the accuracy over the entire set is computed as:

$$cfaccuracy = \frac{100}{M} \sum_i^M \frac{\sum_k^{R_i} \delta_{i,k} h(k)}{\sum_k^{K_i} h(k)} \qquad (35)$$

If the kth item on the ith recommendation list is in the measurement set, then $\delta_{i,k} = 1$, otherwise it is equal to 0. The function $h(k)$ is defined as:

$$h(k) = 2^{\frac{-k}{b}} \qquad (36)$$

where $b$ can be viewed as the "half-life" of $h(k)$, that is $h(k)$ will equal 0.5 when $k = b$. We use a value of 5 for $b$.

As in (Breese et al. 1998), we tested our algorithm on four experiments. For the first three experiments we gave the network 2, 5 and 10 web pages from each test case and asked it to predict the remainder. For the fourth we gave the network all but 1 web page visited by the user and asked it to predict the final web page. For the experiments given 2, 5 and 10 web pages, if fewer than that many pages were in the test case, then the test case wasn't used. At least 2 web pages needed to be present in the test case for use in the all but 1 experiment. Thus each experiment used a different number of test cases. Within our algorithm, Unbounded Rényi Quadratic Entropy (URQE), the confidence coefficient $r$ is set to 5.

For comparison we've supplied results from 5 other algorithms: Bayesian Networks (BN), Correlation technique (CR+) that uses inverse user frequency, default voting and case amplification extensions, Vector Similarity (VSIM) method with an inverse user frequency transformation and Bayesian Clustering (BC). The results for BN, CR+, VSIM and BC are supplied by (Breese et al. 1998). The baseline results are found using the prior probabilities for each web page, i. e. whatever web pages are most visited overall are always chosen regardless of the data given to the network. The value $RD$ is the required difference between two values to be deemed significantly different at the 90% confidence level (Breese et al. 1998).

The results in table 2 show our algorithm, URQE, producing better results in the given 2, 5 and 10 experiments. However, the results are not significantly better. In the all but 1 experiment the URQE outperforms all methods except BN.

| Algorithm | Given 2 | Given 5 | Given 10 | All But 1 |
|---|---|---|---|---|
| URQE | **61.07** | **60.20** | **55.58** | 64.61 |
| BN | 59.95 | **59.84** | **53.92** | **66.69** |
| CR+ | **60.64** | 57.89 | **51.47** | 63.59 |
| VSIM | 59.22 | 56.13 | 49.33 | 61.70 |
| BC | 57.03 | 54.83 | 47.83 | 59.42 |
| Baseline | 49.14 | 46.91 | 41.14 | 49.77 |
| $RD$ | 0.91 | 1.82 | 4.49 | 0.93 |

Table 2: Results for the MS Web data set. The higher the score the better the results. $RD$ is the required difference between scores to be deemed statistically significant. Scores in boldface are within the required difference of the highest score.

| Given 2 | Given 5 | Given 10 |
|---|---|---|
| 9,090 | 5,495 | 2,857 |

Table 3: Recommendations per second for the MS Web data set, given 2, 5 and 10 ratings. All tests are done on a 1 GHz Pentium running Windows 2000.

While producing some of the most accurate results, the URQE is efficient. The URQE is capable of producing between 2,000 and 9,000 queries per second while taking only 5 seconds for learning on a 1GHz Pentium PC, table 3.

### 5.2 Movie Ratings

Our second set of tests involves a database of movie ratings. The database is from the EachMovie collaborative filtering site run by Digital Equipment Research Center from 1995 to 1997. For more information visit *http://research.compaq.com/SRC/eachmovie/*. Each user was asked to rank movies on a 0 to 5 scale. Out of a total of 1,623 movies, each user ranked on average 46.4 movies with a median at 26. There were 4,119 total users in the test set and 5,000 users in the training set. Once again we tested the algorithm on four tasks. The first three gave 2, 5 and 10 movie ratings to the network and the network was asked to predict the remaining ratings given by the user. Our forth task provided the network with all the movie ratings except one and was asked to predict the remaining rating. In all cases, the movie ratings given to the network were chosen randomly from the list of rated movies. In the given 2, 5 and 10 tasks if a user provided fewer than the respective number of ratings, the user was not included in testing. The all but 1 task only included users with at least 2 ratings.

We computed errors based on the absolute deviation between the predicted rating and that given by the user. Once again we provide results from several algorithms provided courtesy of (Breese et al. 1998). More details on the implementation of CR, BC, BN and VSIM can be found in (Breese et al. 1998). The baseline results use the average rating given to the movie by the users.

| Algorithm | Given 2 | Given 5 | Given 10 | All But 1 |
|---|---|---|---|---|
| URQE | **1.059** | **1.014** | **0.982** | **0.928** |
| CR | 1.257 | 1.139 | 1.069 | 0.994 |
| BC | 1.127 | 1.144 | 1.138 | 1.103 |
| BN | 1.143 | 1.154 | 1.139 | 1.066 |
| VSIM | 2.113 | 2.177 | 2.235 | 2.136 |
| Baseline | 1.106 | 1.105 | 1.103 | 1.133 |
| $RD$ | 0.022 | 0.023 | 0.025 | 0.043 |

Table 4: Results for the EachMovie data set. Absolute deviation from the true user ratings. Lower scores indicate better results. $RD$ is the required difference to be deemed statistically significant. Scores in boldface are within the required difference of the lowest score.

| Given 2 | Given 5 | Given 10 |
|---|---|---|
| 1,852 | 1,010 | 581 |

Table 5: Recommendations per second for the EachMovie data set, given 2, 5 and 10 ratings. All tests are done on a 1 GHz Pentium running Windows 2000.

For the our algorithm, URQE, we need to compute the conditional probability matrix **P**. This is made more difficult since the movie ratings are on a scale from 0 to 5 and not binary. To transfer the ratings to a 0 to 1 scale we used the following equation:

$$\tilde{P}(f_i|f_j) = \frac{\sum_k \min(f_i(t_k), f_j(t_k))}{\sum_k f_j(t_k)} \quad (37)$$

If $t_{k,i}$ is unknown for some $i$, then training case $k$ was not used to compute $\tilde{P}(f_i|f_j)$ or $\tilde{P}(f_j|f_i)$ for any $j$. The final movie ratings are computed by multiplying the computed probability estimates by 5. The confidence coefficient $r$ was set to 100 and no complex functions are used.

Surprisingly, the baseline results outperform all the algorithms in the given 2 and 5 experiments except URQE, table 4. Correlation (CR) does outperform the baseline results for the given 10 and all but 1 experiments. URQE produces significantly better results in all of the experiments.

For learning, the URQE took less than a minute on a 1 GHz Pentium running Windows 2000. The learning times for the probabilistic models from (Breese et al. 1998) took up to 8 hours for learning on a 266 MHz Pentium. The correlation based method (CR) was capable of generating 3.2 recommendations per second while the Bayesian network (BN) can generate 12.9 recommendation per second on a 266 MHz Pentium II. In contrast, URQE is capable of generating between 581 and 1,852 recommendations per second, table 5, on a 1 GHz PC.

# 6 Conclusion

Within this paper, we discussed a maximum entropy method for finding approximations to conditional probabilities in large domains. The method uses an entropy measure derived from Rényi's quadratic entropy. Using this measure, our algorithm reduces to solving a set of linear equations that can be efficiently solved. We tested the accuracy of our system on two collaborative filtering data sets with encouraging results. In most cases, our results are better than or equal to the best results while being an order or two in magnitude more efficient.

Besides collaborative filtering, our algorithm may be useful for other applications involving large domains that can be modeled using low-order interactions. Given the efficient method for computing the weights, problems with varying input to output mappings can be handled.

**Acknowledgements**

We would like to acknowledge Microsoft Corporation and Hewlett-Packard Company for use of their databases.